\documentclass[useAMS,usenatbib,fleqn]{mn2e}

\usepackage{txfonts}
\usepackage{epsfig,graphicx,natbib,url,twoopt}
\usepackage{ulem}
\usepackage{times}             

\usepackage[breaklinks=true]{hyperref} 
\usepackage[labelfont=footnotesize,textfont=footnotesize]{caption}  

\bibpunct{(}{)}{;}{a}{}{,}    
\newcommandtwoopt{\citeads}[3][][]{\href{http://adsabs.harvard.edu/abs/#3}%
                                        {\citealp[#1][#2]{#3}}}
\newcommandtwoopt{\citepads}[3][][]{\href{http://adsabs.harvard.edu/abs/#3}%
                                         {\citep[#1][#2]{#3}}}
\newcommandtwoopt{\citetads}[3][][]{\href{http://adsabs.harvard.edu/abs/#3}%
                                         {\citet[#1][#2]{#3}}}
\newcommandtwoopt{\citeyearrads}%
  [3][][]{\href{http://adsabs.harvard.edu/abs/#3}%
                                         {\citeyear[#1][#2]{#3}}}

\defcitealias{pt98}{PT98}
\defcitealias{hz09}{HZ09}
\defcitealias{p13}{P13}

\title[The recent LMC's chemical enrichment history]{Probing the Large Magellanic Cloud's recent chemical enrichment history through its star clusters}

\author[T. Palma et al.]{\parbox[t]{\textwidth}{\vspace{-1cm}
                                 T. Palma$^{1,2,3}$\thanks{e-mail: tpalma@astro.puc.cl},
				 J.J. Clari\'a$^{3.4}$,
				 D. Geisler$^{5}$,
				 L.V. Gramajo$^{3}$,
				 A.V. Ahumada$^{3.4}$
			         }\vspace{0.2cm}\\
$^{1}$Millennium Institute of Astrophysics, Chile.\\
$^{2}$Instituto de Astrof\'isica, Pontificia Universidad Cat\'olica de Chile, Av. Vicu\~na Mackenna 4860, 782-0436 Macul, Santiago, Chile.\\
$^{3}$Observatorio Astron\'omico de C\'ordoba, Universidad Nacional de C\'ordoba, Laprida 854, X5000BGR, C\'ordoba, Argentina.\\
$^{4}$Consejo Nacional de Investigaciones Cient\'ificas y T\'ecnicas (CONICET), Argentina.\\
$^{5}$Departamento de Astronom\'ia, Universidad de Concepci\'on, Casilla 160-C, Concepci\'on, Chile.\\
}

\begin{document}

\date{Accepted XXX. Received XXX; in original form XXX}

\pagerange{\pageref{firstpage}--\pageref{lastpage}} \pubyear{XXXX}

\maketitle

\label{firstpage}


\begin{abstract}
We present Washington system colour-magnitude diagrams (CMDs) for 17 practically unstudied star clusters located in the bar as well as in the inner disc and outer regions of the Large Magellanic Cloud (LMC). Cluster sizes were estimated from star counts distributed throughout the entire observed fields. Based on the best fits of theoretical isochrones to the cleaned $(C-T_1,T_1)$ CMDs, as well as on the $\delta T_1$ parameter and the standard giant branch method, we derive ages and metallicities for the cluster sample. Four objects are found to be intermediate-age clusters (1.8-2.5 Gyr), with [Fe/H] ranging from -0.66 to -0.84. With the exception of SL\,263, a very young cluster ($\sim$ 16 Myr), the remaining 12 objects are aged between 0.32 and 0.89\,Gyr, with their [Fe/H] values ranging from -0.19 to -0.50. We combined our results with those for other 231 clusters studied in a similar way using the Washington system. The resulting age-metallicity relationship shows a significant dispersion in metallicities, whatever age is considered. Although there is a clear tendency for the younger clusters to be more metal-rich than the intermediate ones, we believe that none of the chemical evolution models currently available in the literature reasonably well represents the recent chemical enrichment processes in the LMC clusters. The present sample of 17 clusters is part of our ongoing project of generating a database of LMC clusters homogeneously studied using the Washington photometric system and applying the same analysis procedure.
\end{abstract}


\section{Introduction}
\label{sec:introduction}

One of the nearest galaxies to the Milky Way is the Large Magellanic Cloud (LMC). It is one of only a few galaxies in which the star formation history (SFH) of even the oldest stars can be traced from studying resolved stellar populations. On account of the LMC's proximity and the wealth and variety of its star clusters, they are of fundamental importance for several reasons. In particular, the large number of young and intermediate-age clusters (IACs) in this galaxy makes it easier to understand the chemical enrichment and SFH of the LMC as a whole \citep[e.g.][]{baum13,rich01}, given the relative ease and accuracy with which ages and abundances can be derived for them. Although the overall estimated number of LMC star clusters is $\sim$ 4200 \citep{h88}, this number may be significantly larger if emission-free associations and objects related to emission nebulae are included \citep{bsdl}. This number increases even more if associations with features of somewhat looser clusters and newly formed ones are also included \citep{b08}. Unfortunately, however, the number of well studied LMC clusters still constitutes a very small percentage of those catalogued. Consequently, detailed investigations of even a few clusters represent a significant improvement in our knowledge of the chemical enrichment history of the LMC.\\

The unusual age distribution of the LMC's clusters was first noted by \citet{vdbh}. \citet{DaCosta} was the first one to draw attention to the existence of a substantial gap of about 4-9 Gyr in the age distribution of the star clusters in this galaxy. There appears to be no corresponding age gap for the field stars. The nature and cause of this pronounced cluster age gap still remains a challenging enigma. Although the LMC includes $\sim$ 15 recognized genuine old clusters with ages $\sim$ 12 Gyr 
\citep{olsen98,dutra99}, as well as a rich population of young and IACs, there is only a single cluster (ESO\,121-SC03) known within this age gap. As emphasized by \citet{olsz96}, this gap in the cluster age distribution also represents an ``abundance gap'' in the sense that the old clusters are all metal-poor ($<$[Fe/H]$>$ $\sim$ —1.5 - -2), while the IACs are all relatively metal-rich \citep{olsz91}, approaching even the current abundance in the LMC ($<$[Fe/H]$>$ $\sim$ 
-0.5). Unfortunately, the age gap does not allow us to use star clusters to examine the chemical evolution and star formation history of the LMC during this long period. However, the determination of ages and metallicities is still very useful to trace the details of the recent chemical evolution and identify potential burst(s) in cluster and star formation that occurred over the last few Gyr in the LMC. \\

\begin{figure}
\includegraphics[width=9cm]{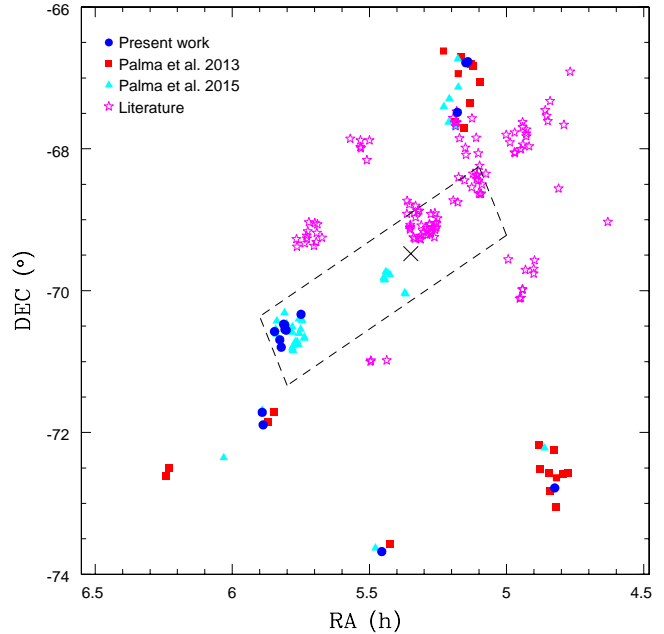}
\caption{ Map of the star clusters studied using the Washington photometric system. The dashed lines delimit the LMC bar region and the cross indicates the geometrical centre \citep{bok}. Blue filled circles, red squares, and magenta triangles represent clusters studied in the current work, in \citetalias{p13}, and in our catalogue in preparation (P15), respectively. Open pink stars stand for clusters studied by other authors using the same technique and analysis procedure.} 
\label{f:fig1}
\end{figure}

Our group has been studying young, intermediate-age and old LMC clusters over the last 10 years using the Washington system \citep[e.g.][]{g03,p03,p11b}. We have chosen to work in this photometric system because of the advantages it offers for this type of study \citep[see, e.g.,][]{g97}. In particular, the combination of the Washington system $C$ and $T_1$ filters is approximately three times more metallicity-sensitive than the $VI$ System \citep{gs99} yet the bands are broad and efficient. We obtained Washington wide-field images for about 21 LMC regions with the ``V\'ictor Blanco'' 4\,m telescope at Cerro Tololo Inter-American Observatory (CTIO, Chile). After a revision of these images, we selected for study a total of 83 star clusters spread out in different regions of the LMC. Ages and metallicities for 23 of them were recently published in \citet[hereafter P13]{p13}. These objects are mostly previously unstudied and are located in the inner disc and outer regions of the LMC. We continue here with our ongoing project presenting results for 17 practically unstudied LMC star clusters, 8 of which (HS\,409, BSDL\,3060, HS\,420, BSDL\,3072, KMHK\,1408, SL\,736, HS\,424, and BSDL\,3123) are projected on the bar. The remaining 9 clusters (SL\,48, KMHK\,575, SL\,263, BSDL\,794, SL\,490, LW\,231, IC\,2140, KMHK\,1504 and SL\,775) are located in the inner disc and outer regions of the LMC (Fig. \ref{f:fig1}). As in \citetalias{p13}, we consider that the inner disc region is limited by a deprojected radius from the LMC centre of $\sim$ 4$^\circ$ \citep{b98}. The position of the cluster NGC\,1928 ($\alpha_{2000}$ = 5$^h$ 20$^m$ 55.9$^s$, $\delta_{2000}$ = -69$^\circ$ 28' 34.9'') was taken as the LMC centre.\\

The positions of our target clusters in relation to the bar and the LMC geometrical  centre are shown in Fig. \ref{f:fig1}. We also include in this figure all the LMC clusters previously studied using the Washington photometric system. The results for the remaining 43 out of the 83 selected clusters will be published in a catalogue of LMC clusters observed in the Washington system (Palma et al., in preparation, hereafter P15).  \\

The cluster sample studied here is presented in Table \ref{t:tab1}, where we list the various star cluster designations from different catalogues, 2000.0 equatorial coordinates, Galactic coordinates, and the cluster core radii given by \citet{b08}. These core radii constitute half of the mean apparent central diameters obtained by computing the average between the major (a) and minor (b) axes. The last two columns of Table \ref{t:tab1} list the cluster radii derived in the current study in arcmin and parsec, respectively (see Sections 3 and 4). We are aware of no other colour-magnitude diagrams (CMDs) for any of these seventeen clusters nor of any other age or metallicity determination. We then aim at deriving ages and metallicities for the present cluster sample. These derived cluster parameters will be combined with those of other LMC clusters previously derived using the same methodology to do our scientific analysis.\\

This paper is organized as follows. The next section briefly describes observations and data reduction. Section 3 focuses on describing the main features of the CMDs and the method applied to minimize field star contamination. In this section, we also describe the procedure followed to estimate cluster radii from the stellar density profiles. Reddenings, ages and metallicities for the seventeen selected clusters are derived in Section 4. A brief analysis and discussion of the results is presented in Section 5. Our main findings are summarized in Section 6.\\


\begin{table*}
 \caption{Observed star clusters in the LMC.}
  \label{t:tab1}
  \begin{tabular}{@{}lccccccc}
  \hline
   Star cluster$^{(a)}$  &  $\alpha_{2000}$ & $\delta_{2000}$ & l & b & r$^{(b)}$ & r$_{cls}$ & r$_{cls}$\\
        & (h m s) & ($\circ$ ' '') & ($\circ$) & ($\circ$) & (') & (') & (pc)\\
 \hline
SL\,48, LW\,68, KMHK\,133 & 04 49 27 & -72 46 54 & 284.859 & -34.698 & 0.45 & 0.90 & 13.1 \\  
KMHK\,575, LOGLE\,139 & 05 08 28 & -66 46 14 & 277.278 & -34.854 & 0.48 & 0.59 & 8.5 \\
SL\,263, LOGLE\,144 & 05 08 54 & -66 47 08 & 277.285 & -34.809 & 0.24 & 0.45 & 6.5 \\
BSDL\,794 & 05 10 46 & -67 29 06 & 278.069 & -34.483 & 0.19 & 0.45 & 6.5 \\
SL\,490, LW\,217, KMHK\,939 & 05 27 18 & -73 40 45 & 284.951 & -31.828 & 0.58 & 0.99 & 14.4 \\
LW\,231, KMHK\,1031 & 05 30 26 & -75 20 57 & 286.813 & -31.270 & 0.45 & 0.52 & 7.5 \\     
IC\,2140, SL\,581, LW\,241,ESO33-24 & 05 33 21 & -75 22 35 & 286.800 & -31.084 & 1.15 & 1.17 & 17.0 \\
HS\,409, KMHK\,1336, LOGLE\,721 & 05 44 57 & -70 19 59 & 280.831 & -31.002 & 0.28 & 0.50 & 7.2 \\
BSDL\,3060 & 05 48 12 & -70 33 24 & 281.061 & -30.710 & 0.37 & 0.59 & 8.5 \\
HS\,420, KMHK\,1403 & 05 48 28 & -70 32 52 & 281.049 & -30.688 & 0.34 & 0.36 & 5.2 \\
BSDL\,3072 & 05 48 33 & -70 28 57 & 280.973 & -30.687 & 0.40 & 0.41 & 5.9 \\
KMHK\,1408 & 05 48 46 & -70 28 23 & 280.959 & -30.670 & 0.55 & 0.72 & 10.5 \\
SL\,736, KMHK\,1420 & 05 49 17 & -70 47 52 & 281.331 & -30.599 & 0.36 & 0.90 & 13.1 \\
HS\,424, KMHK\,1425 & 05 49 36 & -70 41 35 & 281.207 & -30.582 & 0.39 & 0.63 & 9.2 \\
BSDL\,3123 & 05 50 45 & -70 34 34 & 281.063 & -30.497 & 0.23 & 0.32 & 4.6 \\
KMHK\,1504 & 05 53 15 & -71 53 32 & 282.563 & -30.191 & 0.32 & 0.63 & 9.2 \\
SL\,775, LW\,327, KMHK\,1506 & 05 53 27 & -71 42 57 & 282.360 & -30.189 & 0.60 & 0.95 & 13.7 \\
 \hline  
\end{tabular}
 
\medskip
$^{(a)}$ Cluster identifications from (SL): \citetads{sl}; (LW): \citetads{lw}; (KMHK): \citetads{kmhk}; 
(LOGLE): \citetads{logle98,logle99}; (BSDL): \citetads{bsdl}; (HS): \citetads{hs}.
$^{(b)}$ Taken from \citetads{b08}.
\end{table*}


\section{CCD Washington photometric observations and reductions}  

The observations were carried out with the ``V\'ictor Blanco'' 4\,m telescope at CTIO during the nights of 2000 December 29 and 30. As described in \citetalias{p13}, Washington wide-field images of about 21 LMC regions were taken with the MOSAIC II camera, which consists of an 8K$\times$8K CCD detector array. The scale on the MOSAIC wide-field camera is 0.27''/pix, resulting in a 36'$\times$36' field of view on the sky.  To keep consistency with our previous studies, we used the Washington $C$ \citep{canterna} and Kron-Cousins $R$ filters. The latter has a a very similar wavelength coverage but significantly higher throughput as compared with the standard Washington $T_{1}$ filter so that $R$ magnitudes can be accurately transformed to yield $T_1$ magnitudes \citep{g96}. In particular, this filter combination allows us to derive accurate metallicities based on the standard giant branch technique outlined in \citet{gs99}. From here onwards, we will use interchangeably the designations $R_{KC}$ or $T_1$. \\

Exposure times were 450\,s and 150\,s for $C$ and $R_{KC}$ respectively, while airmasses range between 1.29 and 1.59. The seeing was typically 1''-1.5'' during the observing nights. Figs. \ref{f:fig2}-\ref{f:fig18} (upper left-hand panels) show schematic finding charts of the observed cluster fields built with all measured stars in the $T_1$-band. Observational setup, data reduction procedure, stellar point spread function photometry, and transformation to the standard system follow the same prescriptions described in detail in \citetalias{p13}. The transformation in particular is
the same since the data were obtained during the same nights as the data reduced in \citetalias{p13}. The final data collected for each cluster consists of a running star number, the CCD x and y coordinates, the derived T$_1$ magnitudes and $C-T_1$ colours, and the photometric errors $\sigma T_1$ and $\sigma(C-T_1)$. Only a portion of the Washington data obtained for one of the observed clusters (SL\,48) is shown here (see Table \ref{t:tab2}) for guidance regarding their form and content. The whole Washington data for the current studied clusters can be obtained as supplementary material on the on-line version of the journal. \\

As in \citetalias{p13} (see their Fig. 2), the mean magnitude and colour errors for stars brighter than $T_1$ = 20 lie in the range $<\sigma(T_1)>$ = 0.010-0.015 and $<\sigma(C-T_1)>$ = 0.010-0.020; for stars with $T_1$ = 20-22, $<\sigma(T_1)> \le$ 0.05 and $<\sigma(C-T_1)> \le$ 0.06; and for stars with $T_1$ = 22-23, $<\sigma(T_1)> \le$ 0.12 and $<\sigma(C-T_1)> \le$ 0.15. Therefore, the depth and quality of our photometry enabled us to detect and measure the main sequence turnoff (MSTO) for all of these young or intermediate age clusters, which was used in our age estimates.\\

\begin{table}
 \caption{CCD $CT_1$ data of all stars in the field of SL\,48. }
 \label{t:tab2}
 \centering
 \begin{tabular}{ccccccc}
  \hline 
   ID  &  x & y & $T_1$ & $\sigma T_1$ & $C-T_1$ & $\sigma(C-T_1)$ \\
        & (px) & (px) & (mag) & (mag) & (mag) & (mag) \\
\hline 
200 & 18.554 & 1648.657 & 20.849 & 0.019 & 0.549 & 0.022 \\
201 & 18.791 & 3742.105 & 20.420 & 0.013 & 0.406 & 0.015 \\
202 & 18.909 & 2694.696 & 22.081 & 0.036 & 1.095 & 0.050 \\
203 & 18.932 & 2949.530 & 18.735 & 0.005 & 1.818 & 0.007 \\
204 & 19.094 & 3961.119 & 22.145 & 0.049 & 0.888 & 0.064 \\
205 & 19.121 & 3569.717 & 18.971 & 0.005 & 1.987 & 0.009 \\
206 & 19.183 & 1947.643 & 20.813 & 0.019 & 1.247 & 0.023 \\
207 & 19.716 & 1358.663 & 20.929 & 0.013 & 0.702 & 0.017 \\
208 & 19.809 & 1879.302 & 21.761 & 0.036 & 0.906 & 0.049 \\
209 & 20.310 & 2435.889 & 21.747 & 0.026 & 1.175 & 0.040 \\
210 & 20.412 & 2673.790 & 18.732 & 0.005 & 1.861 & 0.007 \\
\hline 
\end{tabular}
\end{table}

\section{Colour-magnitude diagrams and cluster radial density profiles}

\begin{figure}
\includegraphics[width=8.5cm]{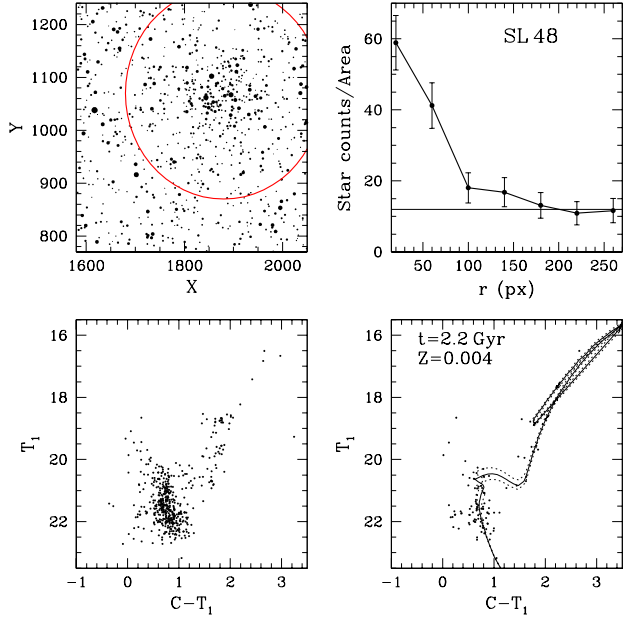}
\caption{Schematic $T_1$-band images of the stars observed in the field of SL\,48 (upper left-hand panel). Cluster stellar density profile, wherein the horizontal line corresponds to the background level far from the cluster (upper right-hand panel). Washington $(T_1,C-T_1)$ CMD for all the stars measured in the cluster field (bottom left-hand panel) and cleaned Washington $(T_1,C-T_1)$ CMD (bottom right-hand panel). The solid line is the isochrone of \citet{bressan} that fits the cluster 
features better, while the dashed lines correspond to the isochrones that best bracket these features.} 
\label{f:fig2}
\end{figure}

\begin{figure}
\includegraphics[width=8.5cm]{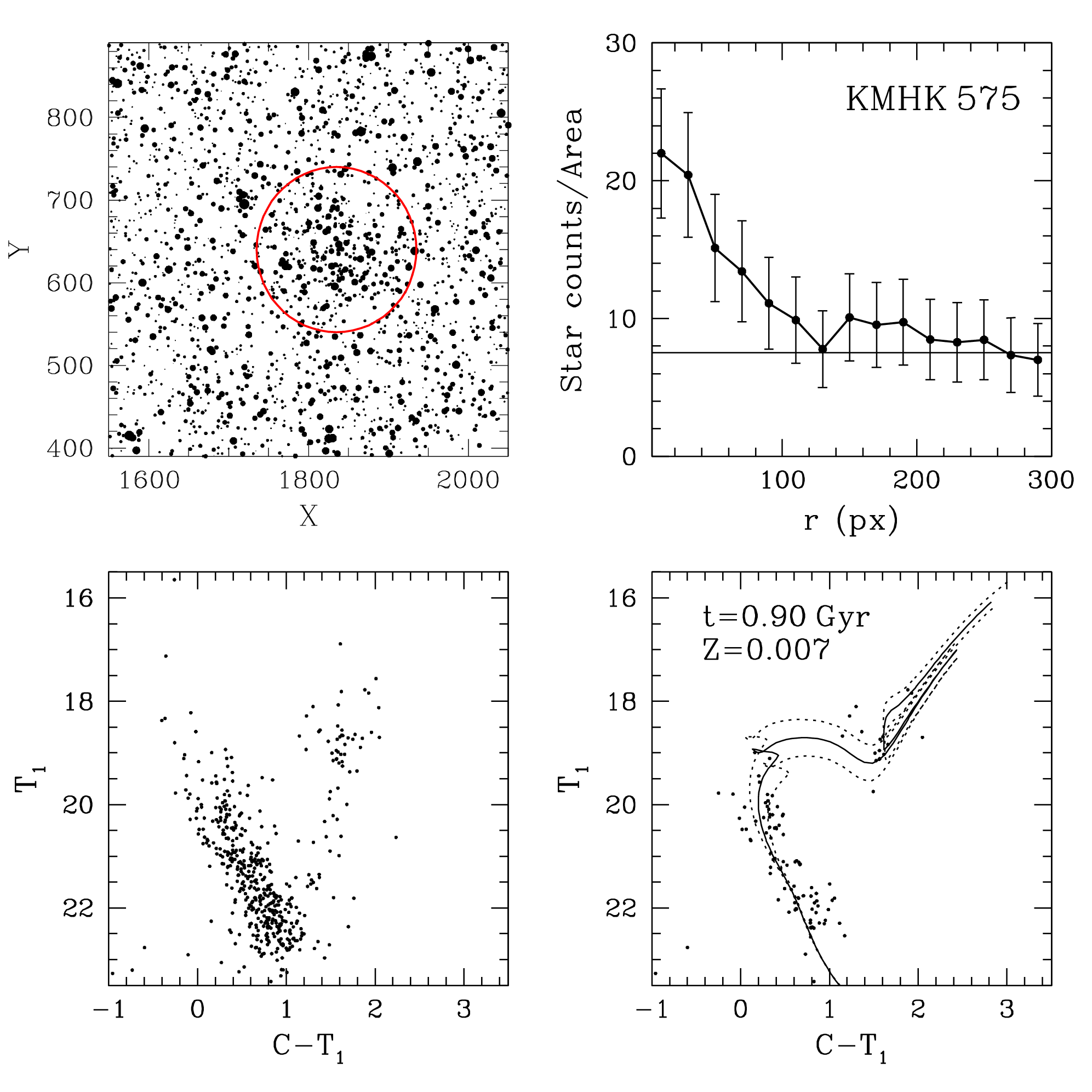}
\caption{Idem Fig. \ref{f:fig2} for KMHK\,575.} 
\label{f:fig3}
\end{figure}

\begin{figure}
\includegraphics[width=8.5cm]{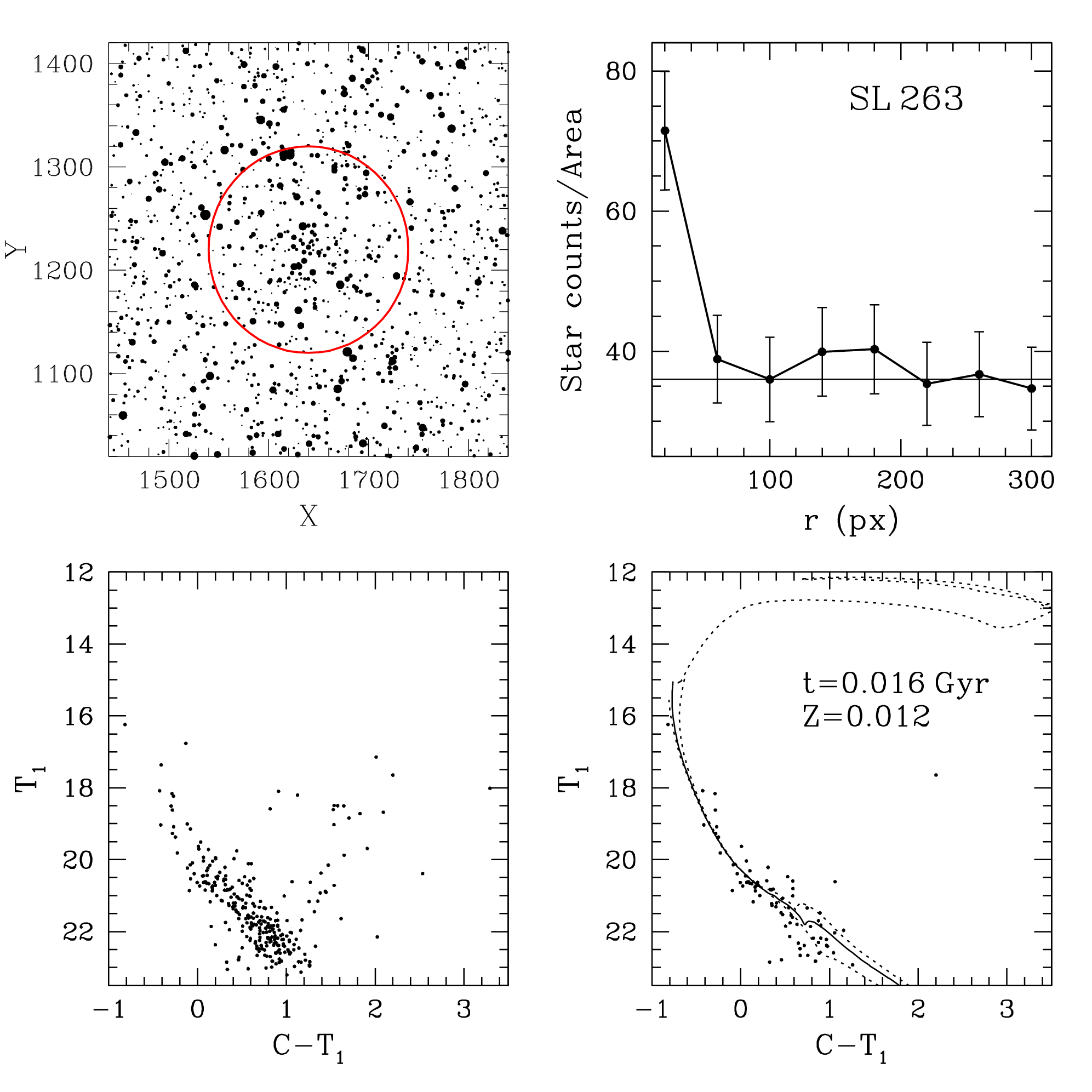}
\caption{Idem Fig. \ref{f:fig2} for SL\,263.} 
\label{f:fig4}
\end{figure}

\begin{figure}
\includegraphics[width=8.5cm]{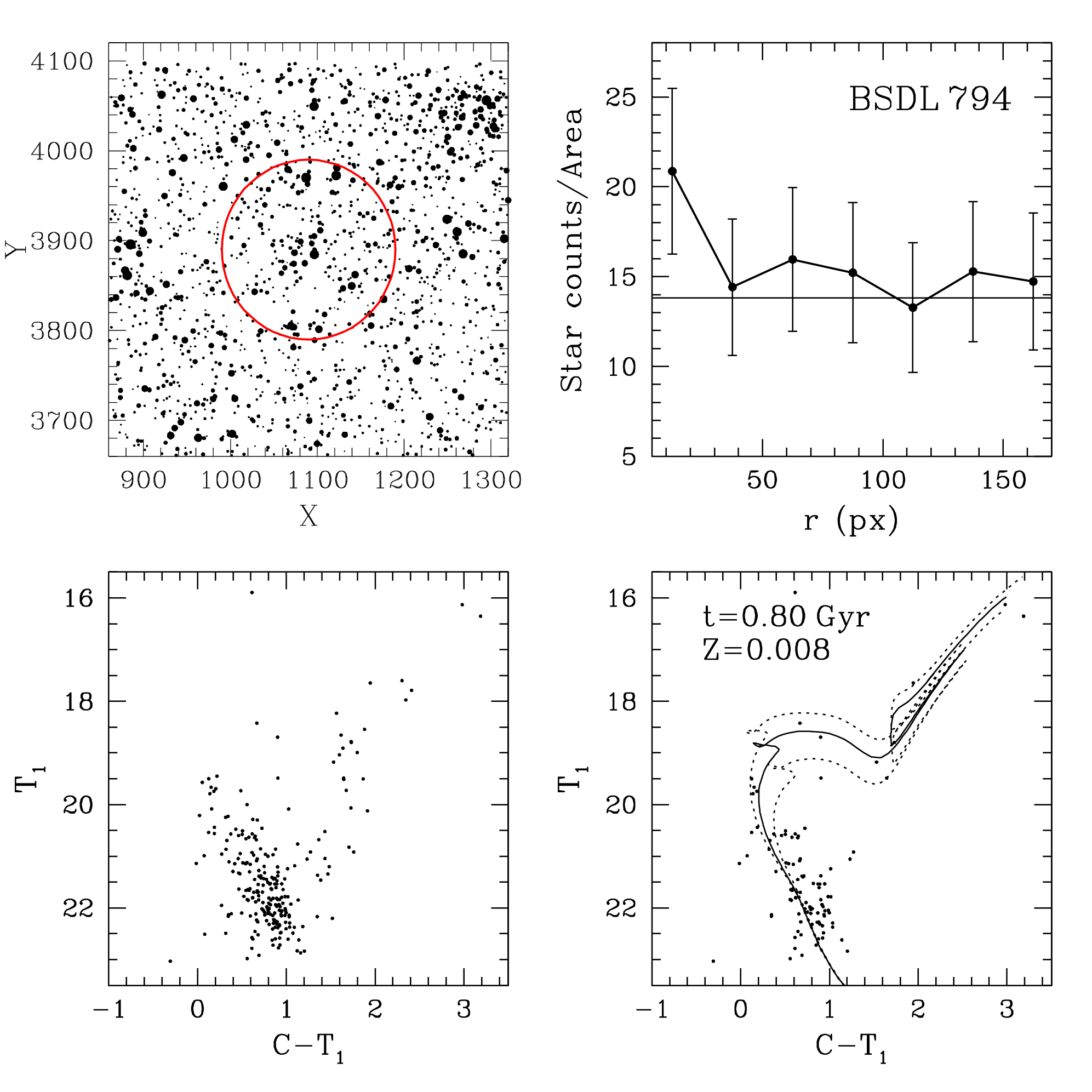}
\caption{Idem Fig. \ref{f:fig2} for BSDL\,794.} 
\label{f:fig5}
\end{figure}

\begin{figure}
\includegraphics[width=8.5cm]{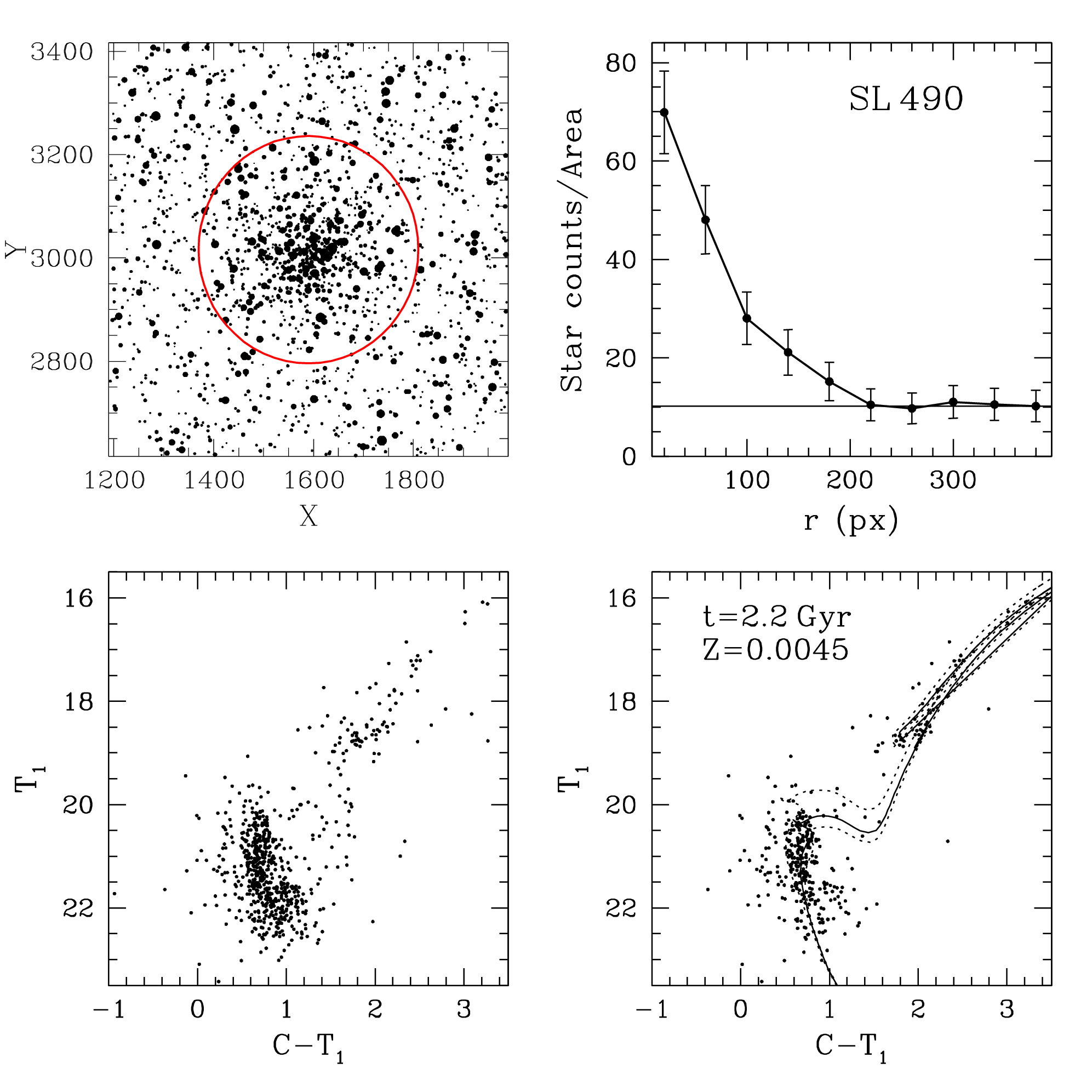}
\caption{Idem Fig. \ref{f:fig2} for SL\,490.} 
\label{f:fig6}
\end{figure}

\begin{figure}
\includegraphics[width=8.5cm]{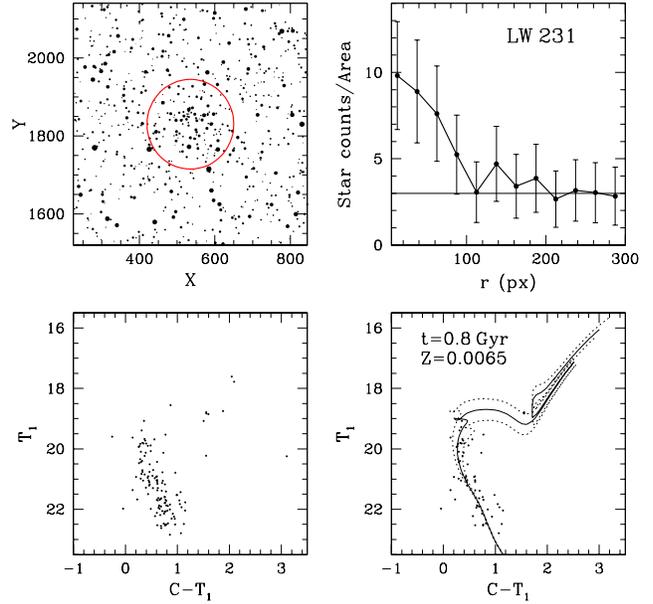}
\caption{Idem Fig. \ref{f:fig2} for LW\,231.} 
\label{f:fig7}
\end{figure}

\begin{figure}
\includegraphics[width=8.5cm]{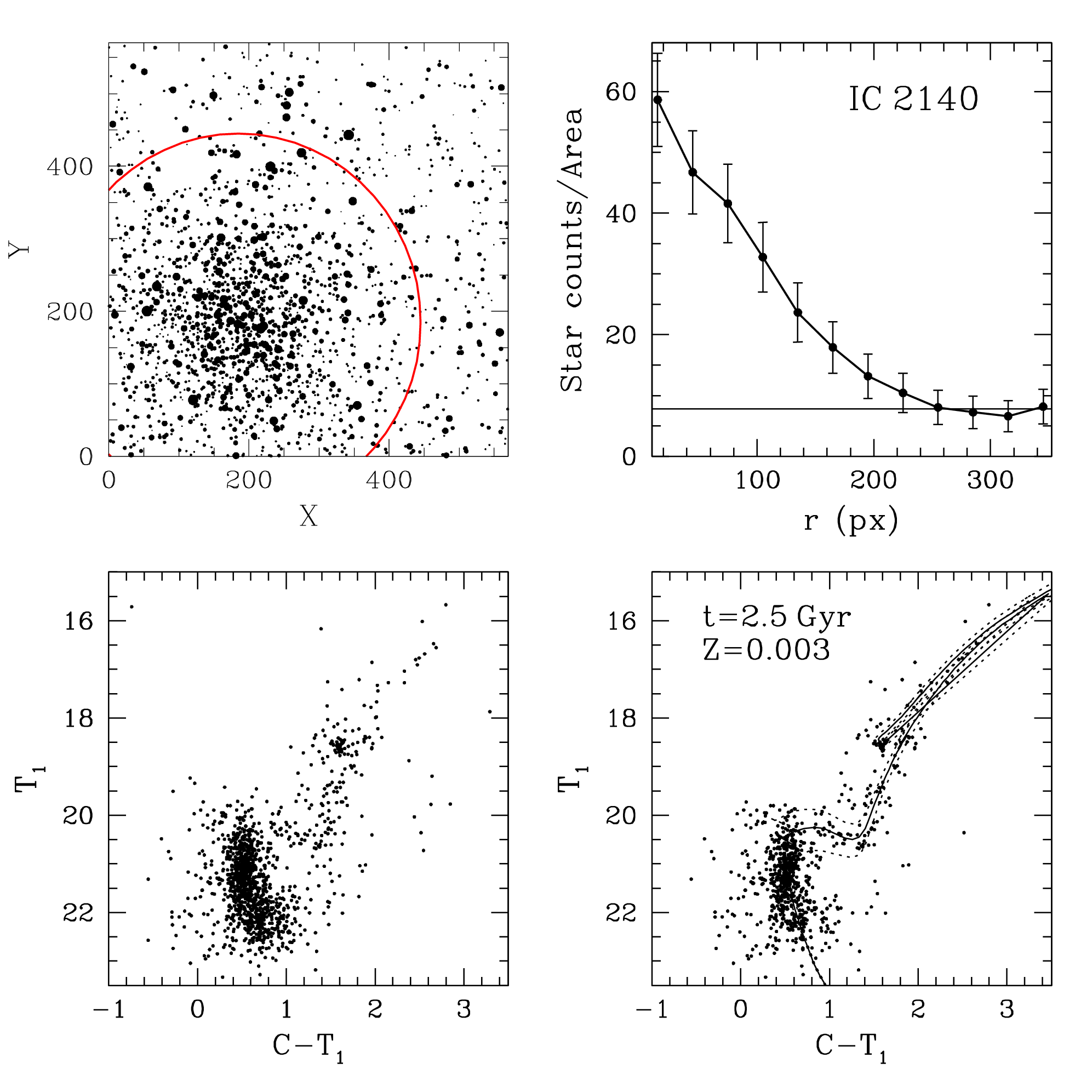}
\caption{Idem Fig. \ref{f:fig2} for IC\,2140.} 
\label{f:fig8}
\end{figure}

\begin{figure}
\includegraphics[width=8.5cm]{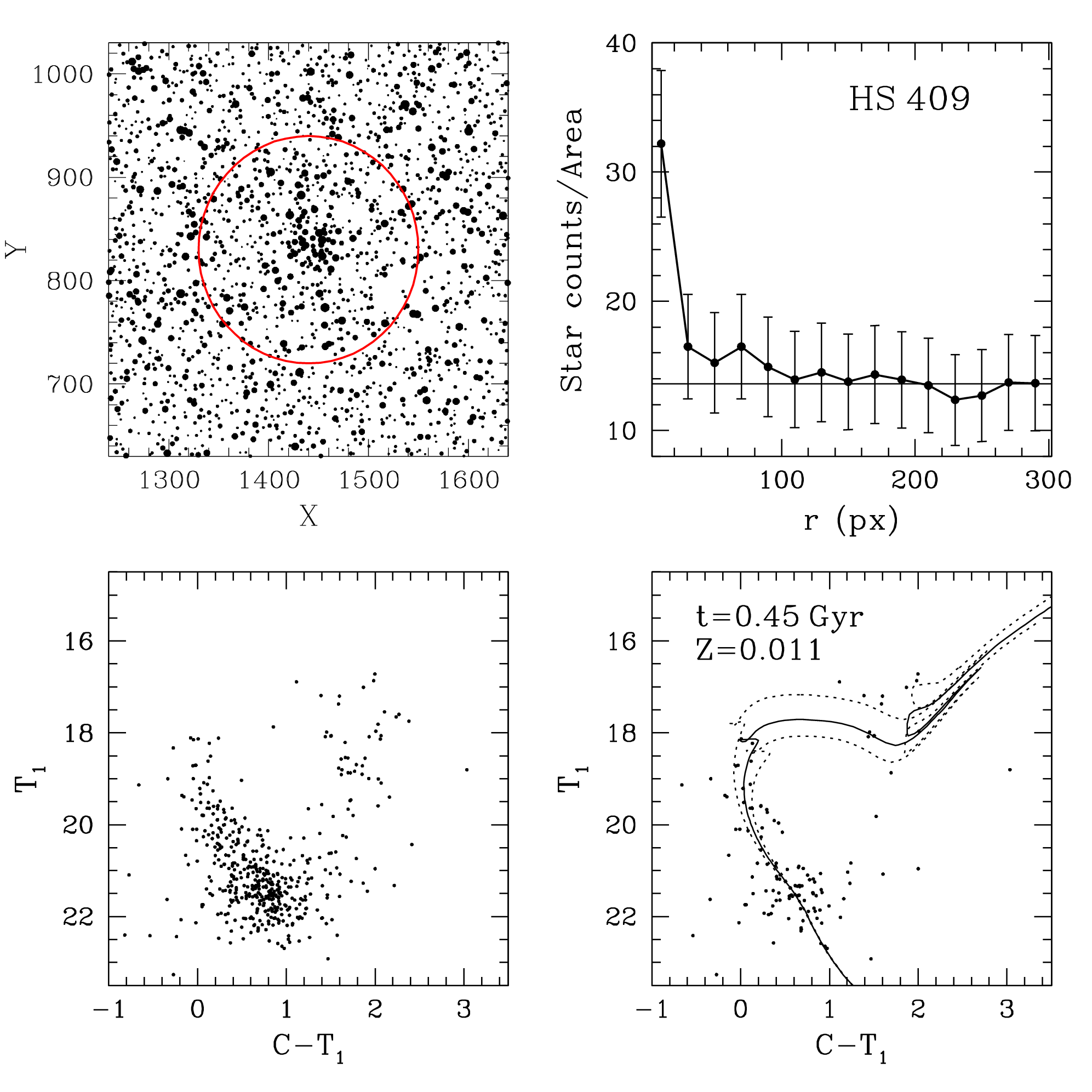}
\caption{Idem Fig. \ref{f:fig2} for HS\,409.} 
\label{f:fig9}
\end{figure}

\begin{figure}
\includegraphics[width=8.5cm]{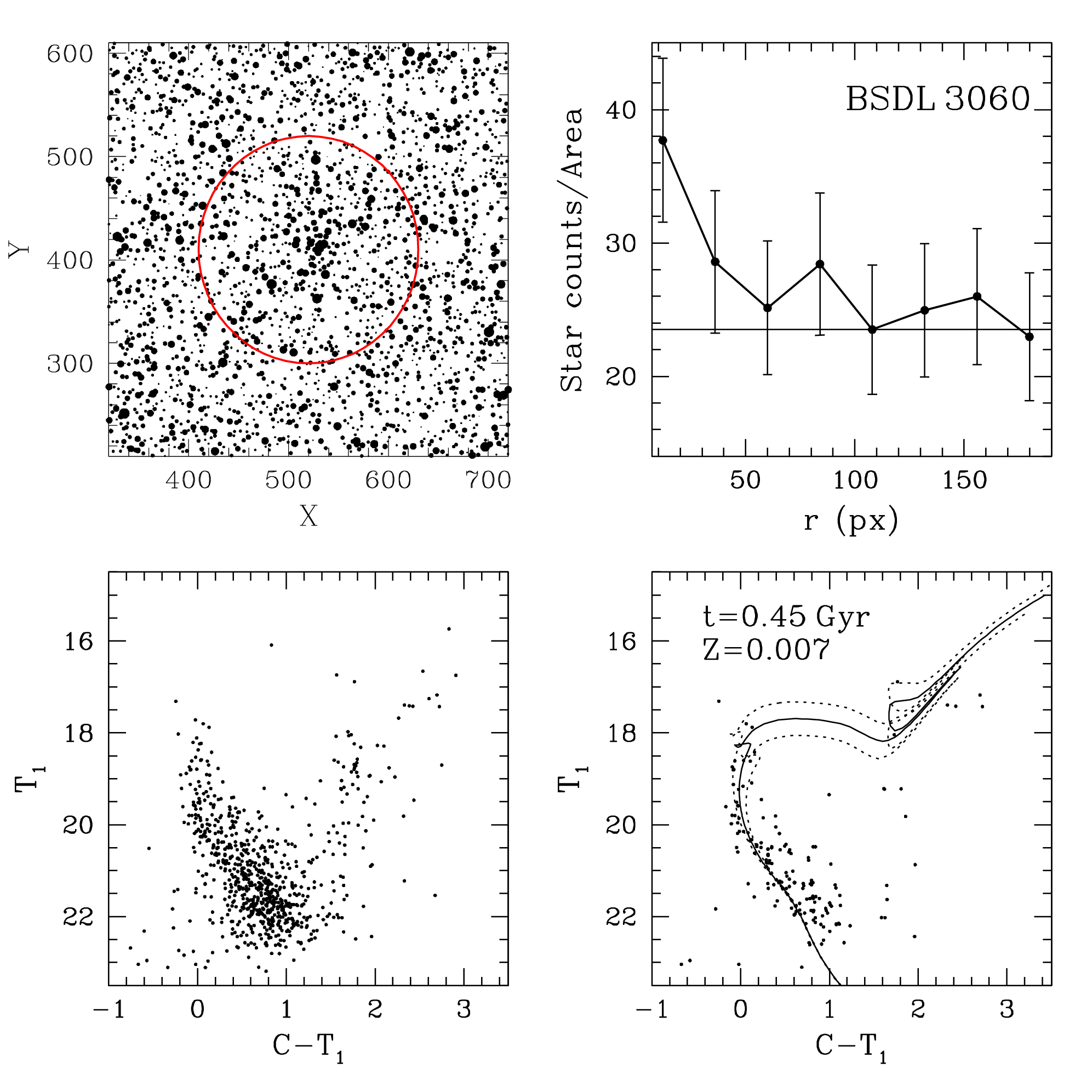}
\caption{Idem Fig. \ref{f:fig2} for BSDL\,3060.} 
\label{f:fig10}
\end{figure}

\begin{figure}
\includegraphics[width=8.5cm]{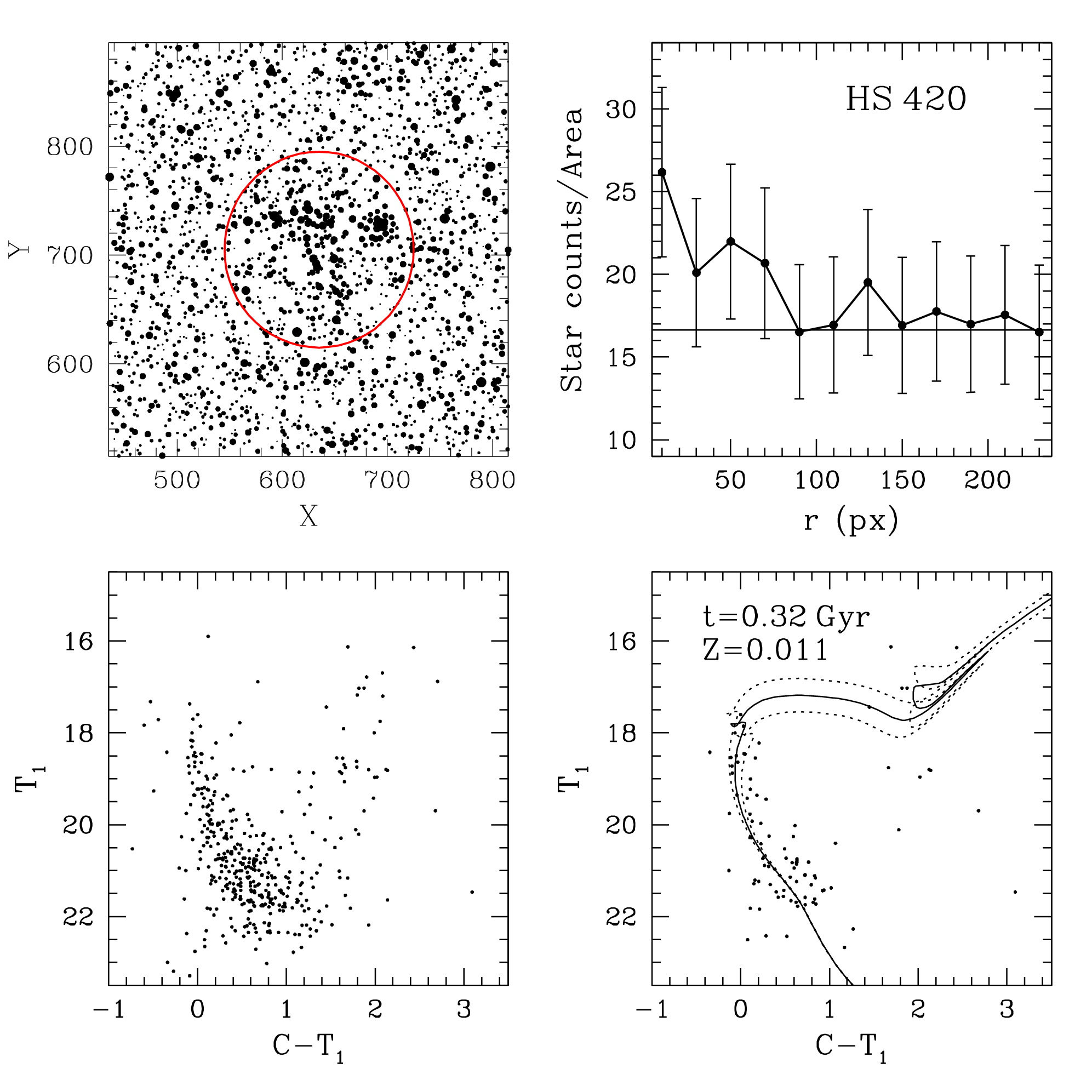}
\caption{Idem Fig. \ref{f:fig2} for HS\,420.} 
\label{f:fig11}
\end{figure}

\begin{figure}
\includegraphics[width=8.5cm]{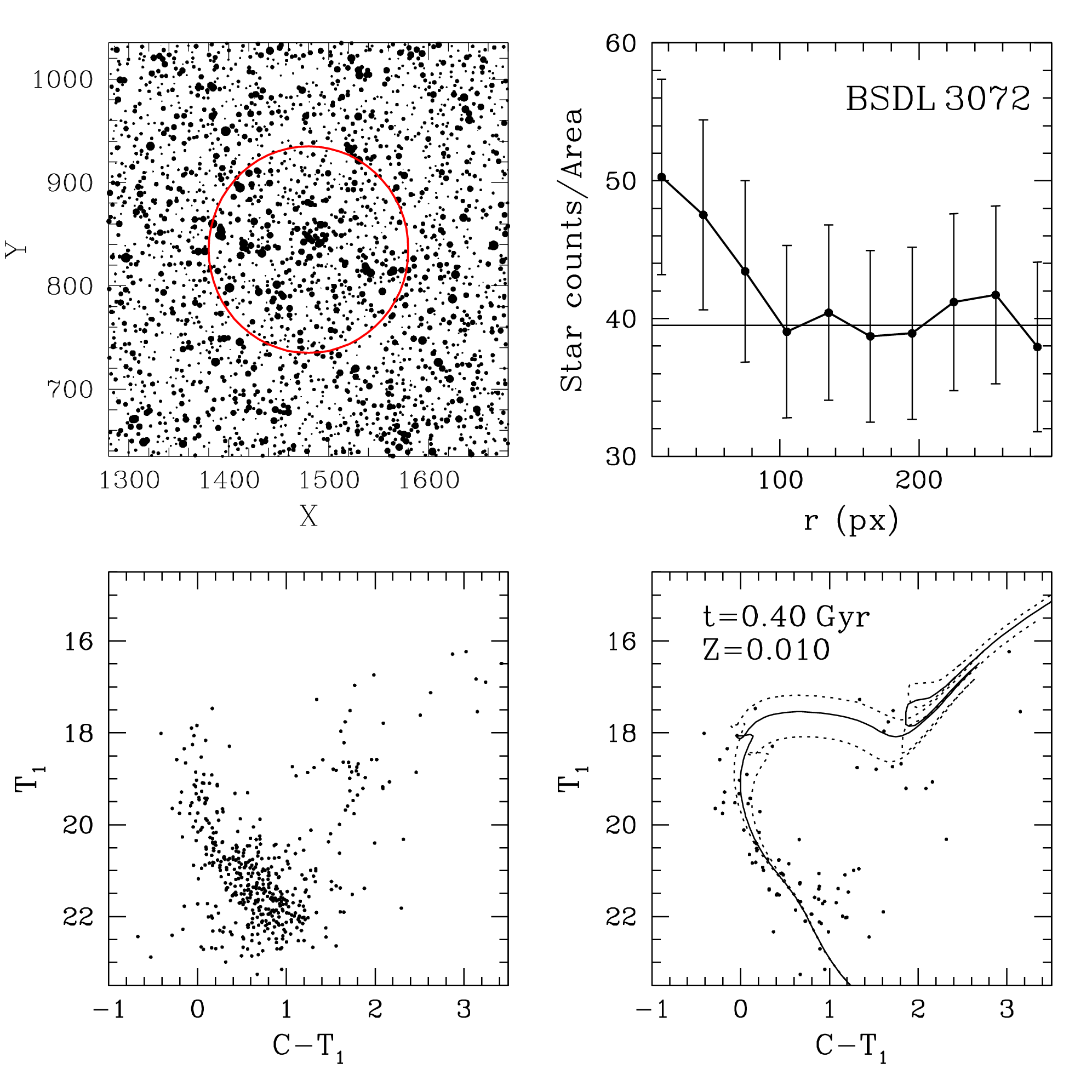}
\caption{Idem Fig. \ref{f:fig2} for BSDL\,3072.} 
\label{f:fig12}
\end{figure}

\begin{figure}
\includegraphics[width=8.5cm]{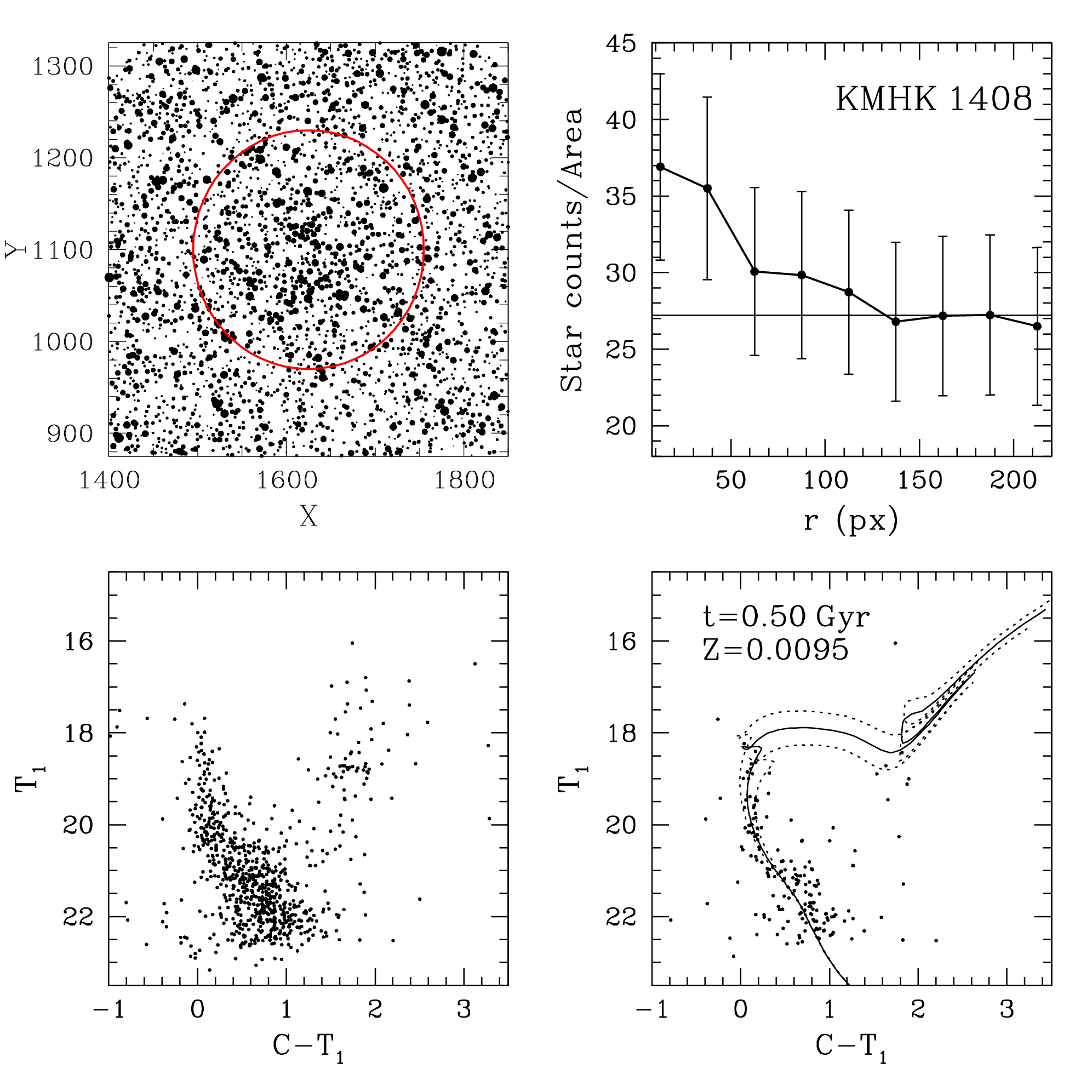}
\caption{Idem Fig. \ref{f:fig2} for KMHK\,1408.} 
\label{f:fig13}
\end{figure}

\begin{figure}
\includegraphics[width=8.5cm]{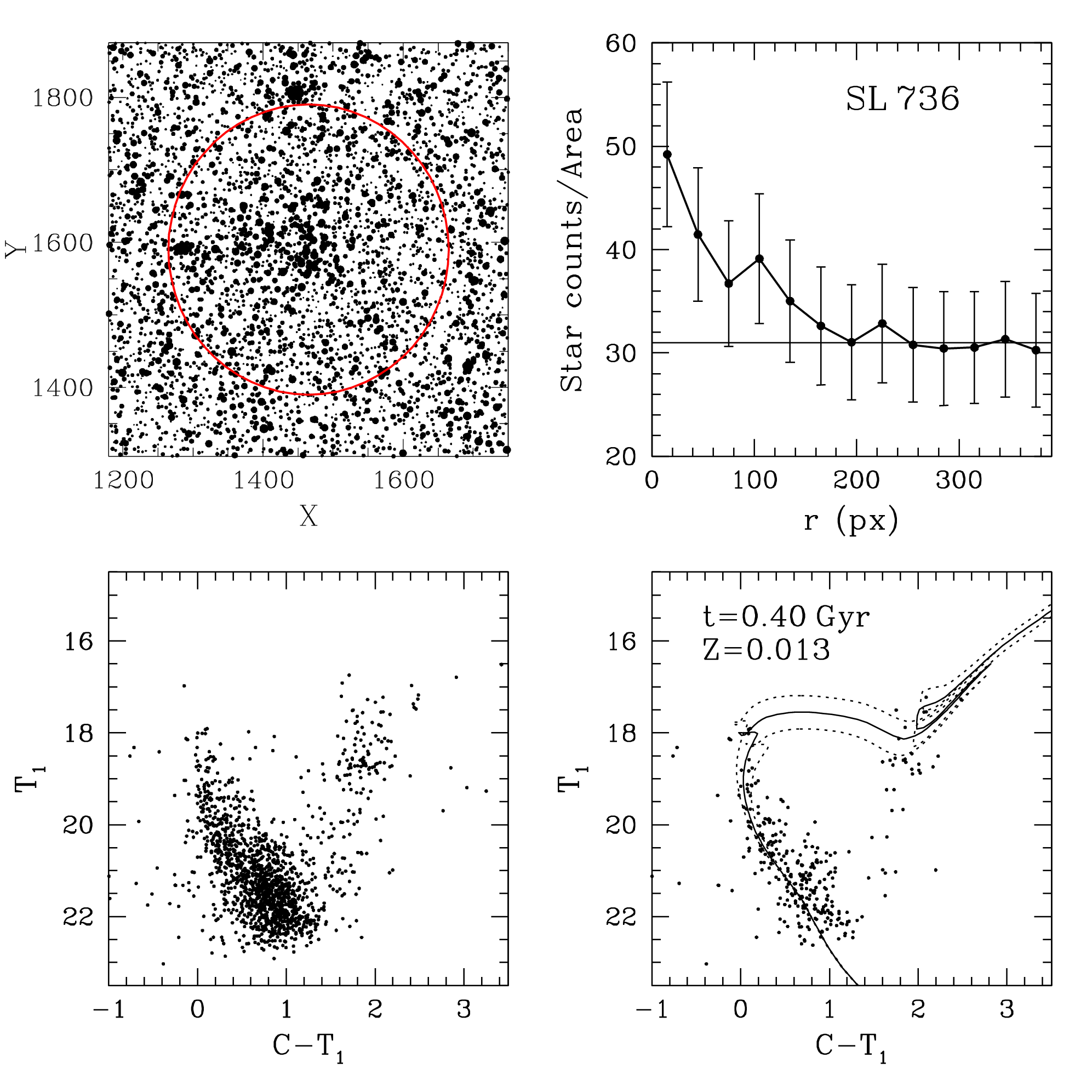}
\caption{Idem Fig. \ref{f:fig2} for SL\,736.} 
\label{f:fig14}
\end{figure}

\begin{figure}
\includegraphics[width=8.5cm]{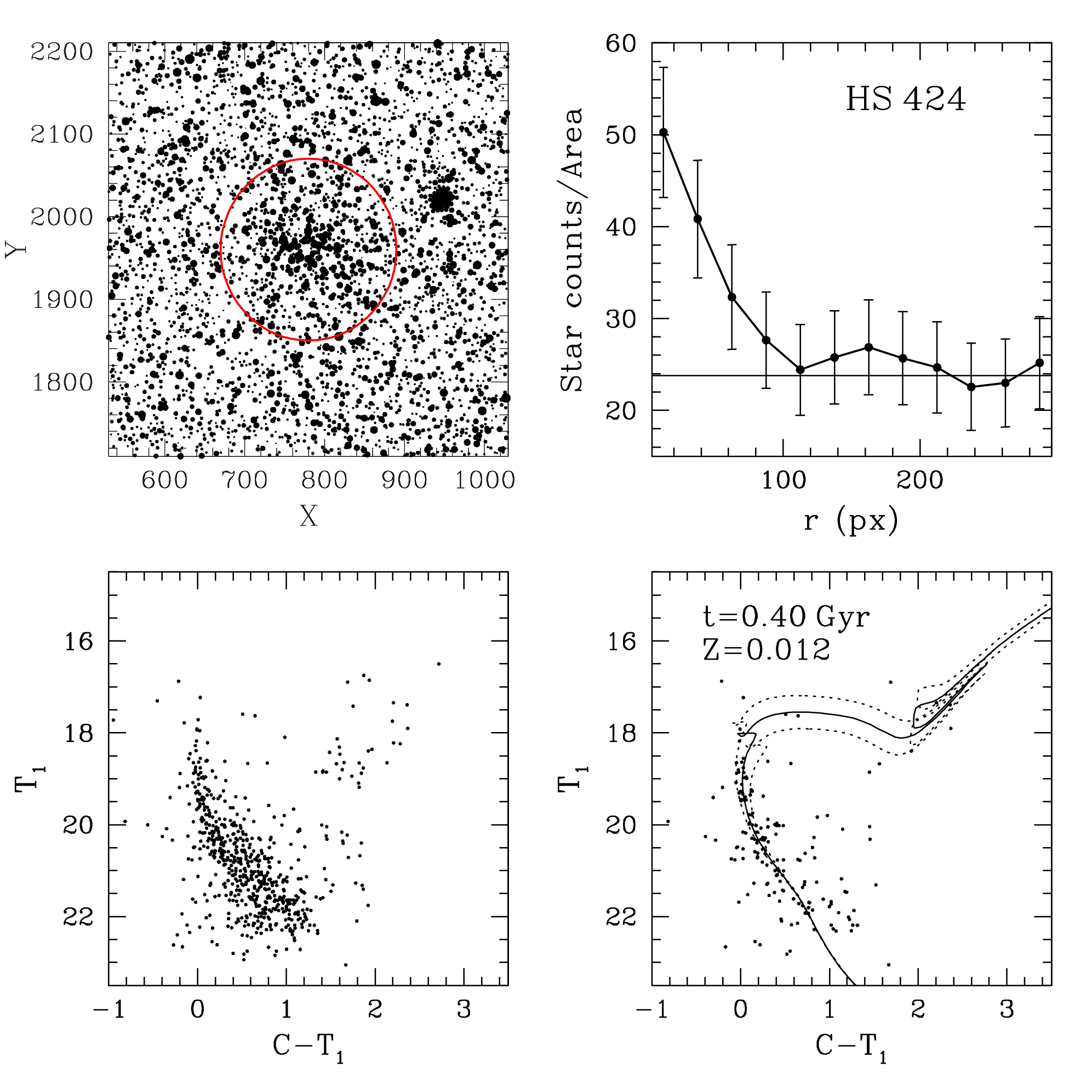}
\caption{Idem Fig. \ref{f:fig2} for HS\,424.} 
\label{f:fig15}
\end{figure}

\begin{figure}
\includegraphics[width=8.5cm]{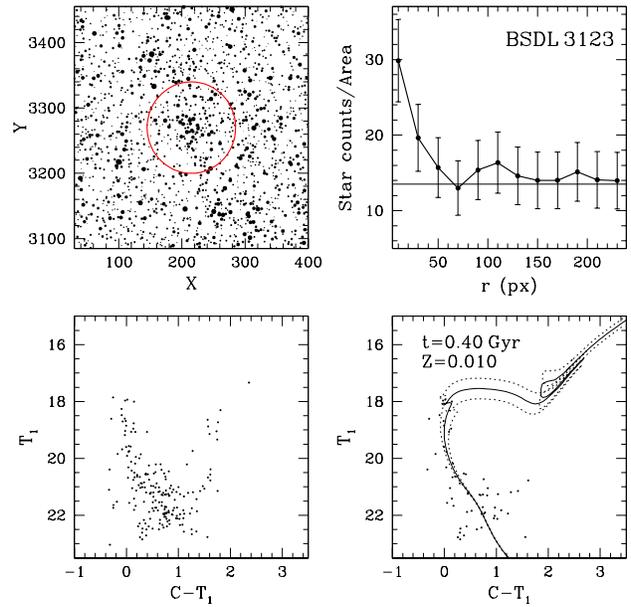}
\caption{Idem Fig. \ref{f:fig2} for BSDL\,3123.} 
\label{f:fig16}
\end{figure}

\begin{figure}
\includegraphics[width=8.5cm]{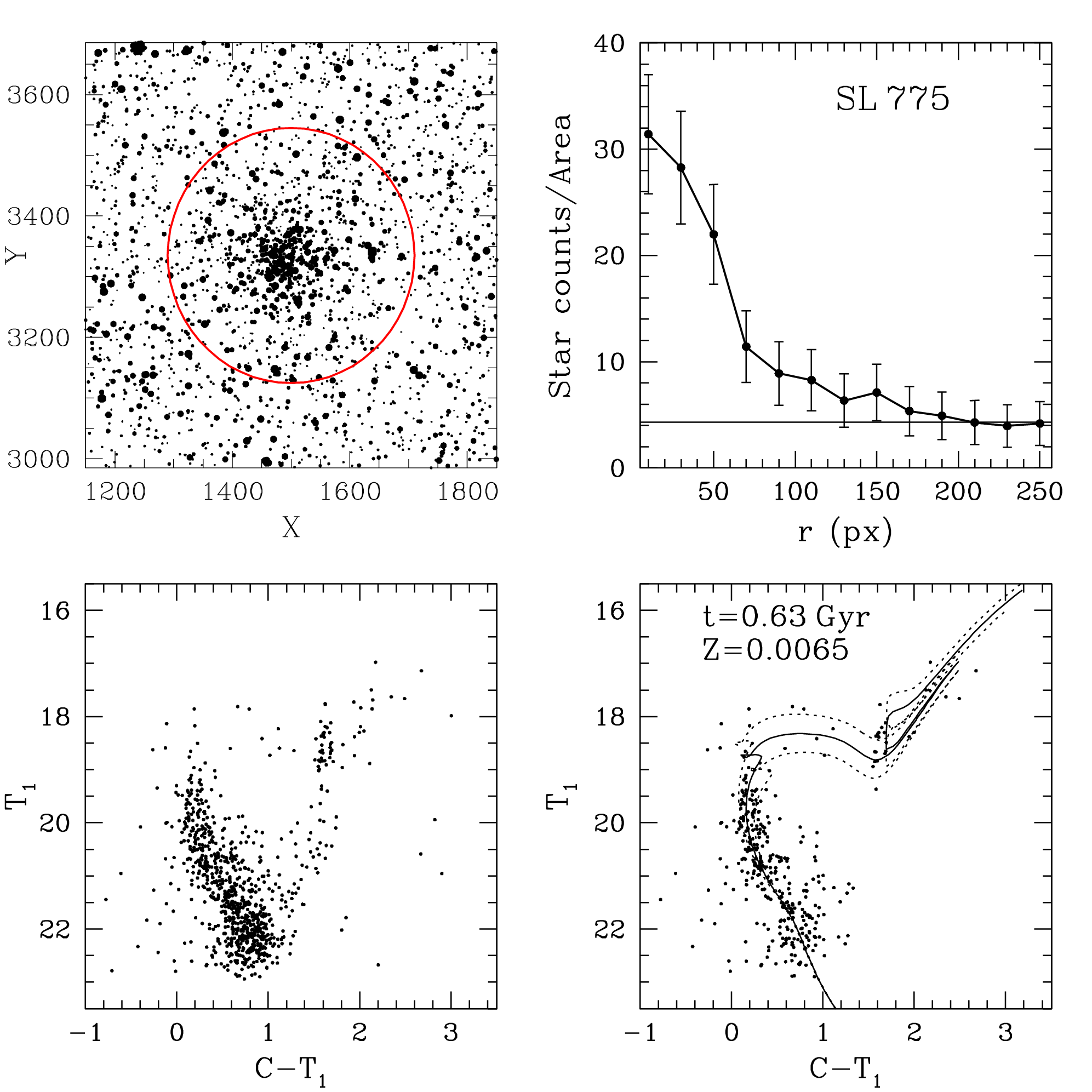}
\caption{Idem Fig. \ref{f:fig2} for SL\,775.} 
\label{f:fig17}
\end{figure}

\begin{figure}
\includegraphics[width=8.5cm]{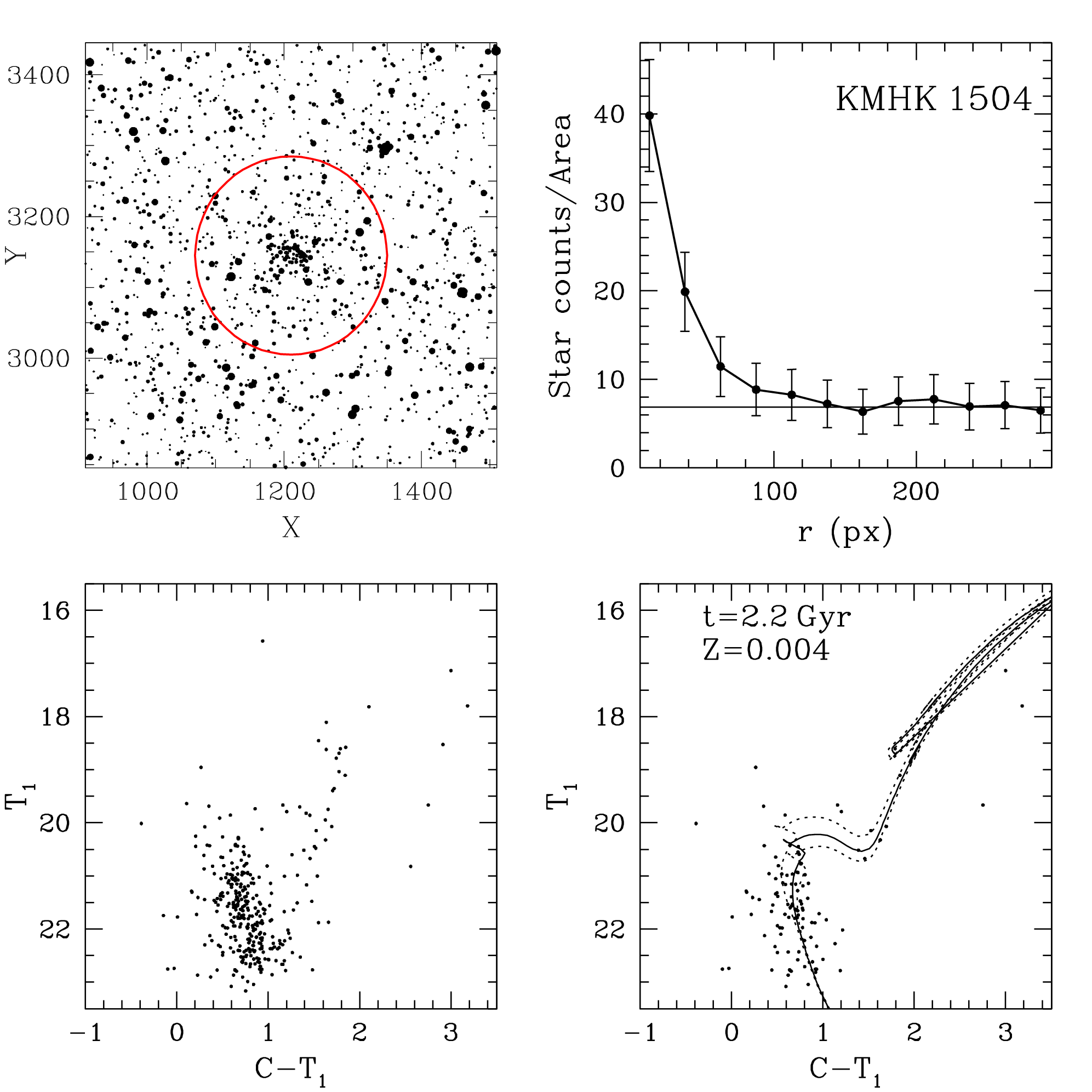}
\caption{Idem Fig. \ref{f:fig2} for KMHK\,1504.} 
\label{f:fig18}
\end{figure}

In the bottom left-hand panels of Figs. \ref{f:fig2}-\ref{f:fig18}, the $(C-T_1,T_1)$ CMDs of all measured stars in the fields of the 17 star clusters are presented. They exhibit very different features depending mainly on age and location in the LMC. Some clusters, particularly the younger ones (SL\,263, for example, Fig. \ref{f:fig4}), present extended main sequences (MSs) with no signs of evolution and no red giant clump (RGC) stars. Other clusters, particularly the older ones of our sample (SL\,48, for example, Fig. \ref{f:fig2}), present less extended MSs, clear signs of evolution, and a variable number of RG stars in the clumps as well as in the giant branches. In addition, they show fairly well developed subgiant branches (SGBs). Of course, a good number of clusters of our sample show CMDs with intermediate features, among others LW\,231 (Fig. \ref{f:fig7}), HS\,409 (Fig. \ref{f:fig9}), and KMHK\,1408 (Fig. \ref{f:fig13}), for example. \\

Before estimating cluster ages and metallicities, we statistically cleaned the cluster CMDs from stars that could potentially belong to the foreground/background fields. We applied a procedure developed by \citet{pb12} and explained therein. The resulting cleaned CMDs of the observed clusters are shown in the bottom right-hand panels of Figs. \ref{f:fig2}-\ref{f:fig18}. These CMDs inevitably contain some residual field star contamination. They allow us, however, to distinguish the MSTOs in all clusters, the RGCs in some of them, as well as MSs extended from approximately two (e.g., IC\,2140, Fig. \ref{f:fig8}) up to seven magnitudes (e.g., SL\,263, Fig. \ref{f:fig4}) downwards. \\

We determined the center of each cluster by building projected histograms in the x and y directions. Using the NGAUSSFIT routine in the STSDAS/IRAF\footnote{IRAF is distributed by the National Optical Astronomy Observatories, which is operated by AURA, Inc., under contract with the National Science Foundation} package, we then fitted Gaussians to these histograms and adopted the centre of these Gaussians as the coordinates of the cluster's centre. Having determined the centre of each cluster and following the procedure described in \citetalias{p13}, we obtained the cluster radial density profile based on star counts carried out in the entire field of each cluster. The resulting  radial density profiles are shown in the upper right-hand panels of Figs. \ref{f:fig2}-\ref{f:fig18}. Column 7 of Table \ref{t:tab1} lists the estimated cluster radius ($r_{cls}$), defined as the distance in arcsec from the cluster's center where the density of stars equals that of the background. The resulting linear radii in parsec calculated assuming that the LMC is located at a heliocentric distance of 50 kpc \citep{s10} are presented in column 8 of Table \ref{t:tab1}. \\

%

\section{Ages and metallicities}

We estimated ages for the cluster sample by fitting theoretical isochrones to the cleaned CMDs. We used isochrones for the Washington system computed by the Geneva  \citep{lesch01} and Padova \citep{bressan} groups. We finally chose to use the \cite{bressan} isochrones because they generally fit the cluster CMDs better. \cite{bressan} set of isochrones are the result of an update of the stellar evolution code used in Padova to compute sets of stellar evolutionary tracks \citep{gir02}. Bressan et al.'s models include much smaller intervals in chemical composition ($Z$) than those used by \cite{gir02}so that more accurate fittings can then be obtained. In general, the different sets of theoretical isochrones both from Geneva and Padova lead to similar results. A detailed analysis of the differences arising when different sets of isochrones are used can be seen in P15 (in preparation).\\

To fit \cite{bressan} isochrones into the cluster CMDs, we first estimated cluster reddening values by interpolating the extinction maps of \citet{bh82}. We decided not to use the full-sky maps from 100-$\mu$ dust emission obtained by \citet{sfd98} for the reasons given in previous papers \citep[see, e.g.,][]{g03}. As shown in column 3 of Table \ref{t:tab4}, the resulting $E(B-V)$ values range between 0.04 and 0.12 mag. Such variations of Galactic reddening are to be expected 
since the distribution of the present clusters covers as much as 16$^{\circ}$ on the sky. \\

We used a true distance modulus for the LMC of $(m-M)_o$ = 18.50 $\pm$ 0.10, as reported by \citet{s10}. The effect that produces the line-of-sight depth of the LMC on the individual cluster distances may be considered negligible. According to \citet{subramanian}, the average depth for the LMC disc is $3.44 \pm 0.16$ kpc. Thus, provided that any cluster of our sample could  be located in front of or behind the LMC main body, the difference in apparent distance modulus could be as large as $\sim$0.3 mag. Since we believe the error involved when adjusting the isochrones to the cluster CMDs is $\sim$0.2-0.3 mag, our assumption of adopting one single value for the distance modulus of all the clusters should not affect our final results substantially. We then selected a set of isochrones computed taking into account overshooting effects, along with the equations $E(C-T_1) = 1.97E(B-V)$ and $M_{T_1} = T_1 + 0.58E(B-V) - (V-M_V)$ given by \citet{gs99}. Next, we superimposed the isochrones on the cleaned cluster CMDs, once they were properly shifted by the corresponding $E(B-V)$ colour excess and LMC apparent distance modulus. We used chemical compositions in the interval 0.003 $\leq$ $Z$ $\leq$ 0.013, equivalent to -0.84 $\leq$ [Fe/H] $\leq$ -0.19 for the isochrone sets in steps of $\Delta$log $t$ = 0.05 dex. The age corresponding to the isochrone that best reproduced the shape and position of the cluster MS, particularly at the MSTO level, 
was adopted as the cluster age. We also took into account the $T_1$ magnitude of the RGC. Note, however, that the theoretically computed bluest stage during the He-burning core phase is redder than the observed RGC in the CMD of IC\,2140 (Fig. \ref{f:fig8}), a behaviour previously detected in other studies of Galactic and Magellanic Cloud clusters \citep[e.g.,][]{c07,p11a}. The age error was estimated considering the isochrones that encompassed those features best. The bottom 
right-hand panels of Figs. \ref{f:fig2}-\ref{f:fig18} as well as columns 5 and 6 of Table \ref{t:tab4} show the results of the isochrone fittings. \\

On the other hand, it is well known that the luminosity of the MSTO depends on the cluster age, but the luminosity of the RGC is almost age independent \citep{cannon}. This is why the $\delta T_1$ parameter, defined as the difference in $T_1$ magnitude between the RGC and the MSTO in the Washington ($T_1$,$C-T_1$) CMD, is a good and reliable parameter to be used as an age indicator. \citet{g97} calibrated $\delta T_1$ as a function of age for IACs, i.e. generally older than 1 Gyr. Since four of the present clusters (SL\,48, SL\,490, IC\,2140 and KMHK\,1504) are clearly IACs, we derived their ages based on the $\delta T_1$ parameter. Their RGCs have an average magnitude $(T_1)_{clump}$ = 18.7 $\pm$ 0.1 mag. The corresponding MSTOs of the cluster sample appear to be more difficult to determine, mainly because of the crowded nature of some cluster fields, intrinsic dispersion and photometric errors at these faint magnitudes. The mean $\delta T_1$ values and their errors were estimated from independent measurements of MSTOs and RGCs made by two different authors (TP and JJC) using lower and upper limits in order to take into account the 
intrinsic dispersion. Such independent measurements showed in general very good agreement. Column 4 of Table \ref{t:tab4} lists the resulting cluster ages computed 
with equation (4) of \citet{g97}. The $\delta T_1$ ages appear systematically lower for these four clusters but the differences are within the errors of the respective techniques. \\

\begin{figure}
\includegraphics[width=8.8cm]{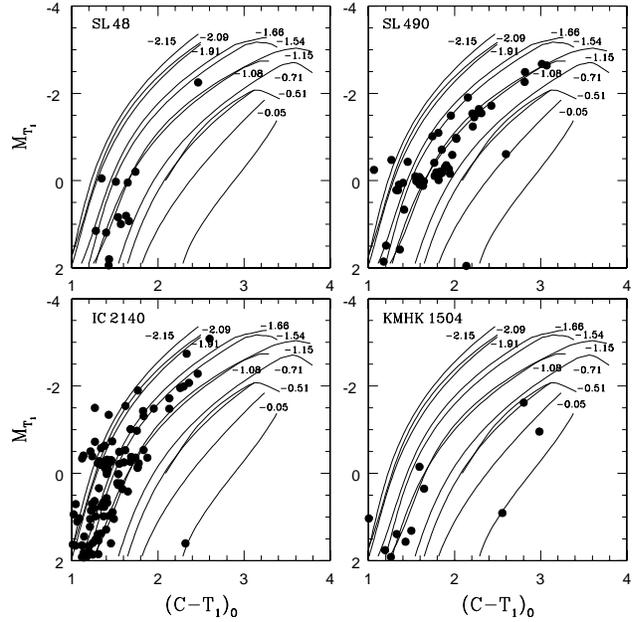}
\caption{Diagram of Washington M$_{T_1}$ vs. $(C-T_1)_0$ for upper RGB stars in four LMC star clusters of the sample, with Standard Giant Branches from \citet{gs99} superimposed. 
An age-dependent correction to the indicated metallicities was applied, as explained in the text.} 
\label{f:fig19}
\end{figure}

We then followed the standard giant branch procedure of entering into the results shown in Figure 4 of \citet{gs99} the absolute $M_{T_1}$ magnitudes and intrinsic $(C-T_1)_o$ colours of upper red giant branch stars for these four clusters to roughly derive their metal abundances [Fe/H] by interpolation (Fig. \ref{f:fig19}). The metallicities herein derived were then corrected for age effects following the prescriptions given in \citet{g03}. The final age-corrected [Fe/H] values are listed in column 7 of Table \ref{t:tab4} and have an associated error of at least
0.2 dex. We found a good agreement between these values and those corresponding to the isochrones that best resemble the cluster CMDs (column 6 of Table \ref{t:tab4}). Note that \citet{gir02} models are computed for [Fe/H] = -0.7, -0.4 and 0.0 dex but not for intermediate metallicity values. However, note that the number of presumed cluster member giants brighter than the clump, where this technique is most sensitive to metallicity, is very small for SL\,48 and KMHK\,1504 so that these metallicity values are especially uncertain. We finally adopt the isochrone metallicity as our official value since that way we have metallicities for all of the clusters obtained with the same procedure.\\

\section{Analysis and discussion}

In order to have a more representative and wider sample of LMC star clusters for a thorough and reliable analysis, we compiled data from the literature and extended our current sample of 17 clusters to a total of 248 clusters studied using the Washington photometric system. The latter represents a unifom and homogeneous sample of LMC star clusters since all of them were observed at CTIO in the Washington photometric system and were then analyzed by applying the same procedure. 
Details on this extended sample such as the cluster fundamental parameters and corresponding references can be seen in P15 (in preparation). 
Notice that our sample of 83 clusters increased by $\sim$ 50\% the number of clusters previously studied using the Washington system.  \\

\begin{table*}
 \caption{Fundamental parameters of LMC clusters}
 \centering
 \begin{tabular}{lcccccc}
 \hline \hline
 Name  &  Deprojected & E(B-V) & $\delta T_1$ Age & Isochrone Age & [Fe/H]$_{isochrone}$ & [Fe/H]$_{SGB}$  \\
       &   Distance ($^{\circ}$) &  & (Gyr)  & (Gyr) &  &  \\
\hline
\vspace{0.15cm}
SL\,48     & 5.1 & 0.12 & 2.1$\pm$0.4 & 2.5$^{+0.3}_{-0.3}$ & -0.72$^{+0.10}_{-0.06}$ & -0.8 \\  
\vspace{0.15cm}
KMHK\,575  & 4.2 & 0.04 & -- & 0.89$^{+0.23}_{-0.19}$ & -0.47$^{+0.11}_{-0.07}$ & -- \\
\vspace{0.15cm}
SL\,263    & 3.6 & 0.04 & -- & 0.016$^{+0.006}_{-0.005}$ & -0.23$^{+0.23}_{-0.17}$ & -- \\
\vspace{0.15cm}
BSDL\,794  & 2.7 & 0.06 & -- & 0.80$^{+0.30}_{-0.17}$ & -0.41$^{+0.09}_{-0.14}$ & -- \\ 
\vspace{0.15cm}    
SL\,490    & 4.8 & 0.11 & 1.8$\pm$0.4 & 2.2$^{+0.3}_{-0.4}$ & -0.66$^{+0.12}_{-0.27}$ & -0.8 \\
\vspace{0.15cm}
LW\,231    & 4.7 & 0.11 & -- & 0.80$^{+0.20}_{-0.17}$ & -0.50$^{+0.14}_{-0.22}$ & -- \\
\vspace{0.15cm}
IC\,2140   & 6.8 & 0.11 & 2.1$\pm$0.4 & 2.5$^{+0.6}_{-0.5}$ & -0.84$^{+0.22}_{-0.18}$ & -1.1 \\
\vspace{0.15cm}
HS\,409    & 2.3 & 0.07 & -- & 0.45$^{+0.11}_{-0.09}$ & -0.27$^{+0.22}_{-0.14}$ & -- \\
\vspace{0.15cm}
BSDL\,3060 & 2.6 & 0.07 & -- & 0.45$^{+0.11}_{-0.10}$ & -0.47$^{+0.31}_{-0.15}$ & -- \\     
\vspace{0.15cm}
HS\,420    & 2.6 & 0.07 & -- & 0.32$^{+0.08}_{-0.07}$ & -0.27$^{+0.17}_{-0.20}$ & --   \\
\vspace{0.15cm}
BSDL\,3072 & 2.6 & 0.07 & -- & 0.40$^{+0.16}_{-0.08}$ & -0.31$^{+0.21}_{-0.31}$ & -- \\
\vspace{0.1cm}
KMHK\,1408 & 2.6 & 0.07 & -- & 0.50$^{+0.13}_{-0.10}$ & -0.33$^{+0.10}_{-0.24}$ & -- \\
\vspace{0.15cm}
SL\,736    & 2.7 & 0.07 & -- & 0.40$^{+0.10}_{-0.08}$ & -0.19$^{+0.17}_{-0.22}$ & -- \\ 
\vspace{0.15cm}
HS\,424    & 2.7 & 0.07 & -- & 0.40$^{+0.10}_{-0.08}$ & -0.23$^{+0.16}_{-0.21}$ & -- \\
\vspace{0.15cm}
BSDL\,3123 & 2.8 & 0.07 & -- & 0.40$^{+0.10}_{-0.08}$ & -0.31$^{+0.24}_{-0.26}$ & -- \\
\vspace{0.15cm}
KMHK\,1504 & 3.6 & 0.12 & 2.1$\pm$0.4 & 2.2$^{+0.3}_{-0.4}$ & -0.72$^{+0.18}_{-0.12}$ & -0.8 \\
\vspace{0.15cm}
SL\,775    & 3.5 & 0.10 & -- & 0.63$^{+0.17}_{-0.13}$ & -0.50$^{+0.23}_{-0.22}$ & --\\
\hline 
\label{t:tab4}
\end{tabular}
\end{table*}

\begin{figure}
\includegraphics[width=9cm]{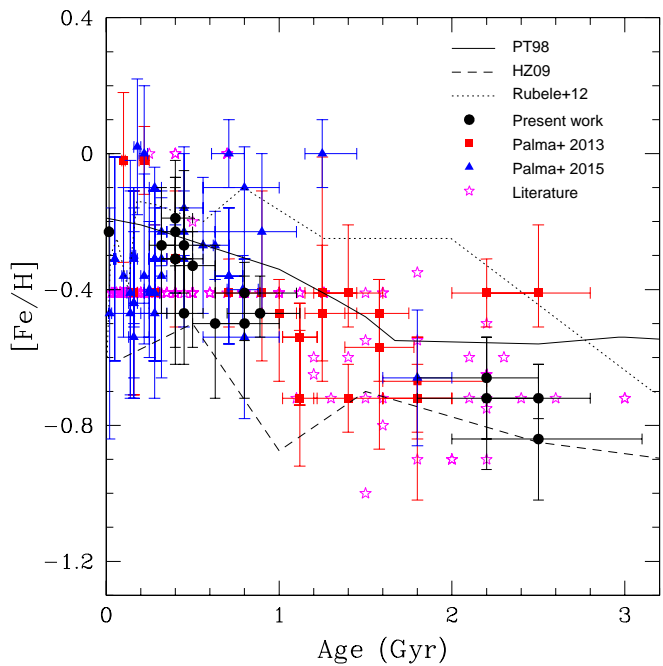}
\caption{Age-metallicity relation for LMC star clusters studied using the Washington photometric system.} 
\label{f:fig20}
\end{figure}

The study of a galaxy's age-metallicity relation (AMR) provides a strong indicator of its chemical evolution history. In this regard, a homogeneous cluster sample represents a more accurate and reliable testimony of the galaxy's chemical evolution history. In Fig. \ref{f:fig20}, we can see the resulting AMR when the full sample of 248 clusters is considered. Symbols and colours are the same as in Fig. \ref{f:fig1}. We overlapped in Fig. \ref{f:fig20} the most widely accepted models used in the literature for the LMC chemical evolution, based on star clusters studies. The \citet[hereafter PT98]{pt98} model is based on a bursting model wherein star formation is assumed to be constant for clusters in the range 1.6 $\leq$ age (Gyr) $\leq$ 3.2 of Fig. \ref{f:fig20} and in which the metallicity increases systematically for clusters younger than 1.6 Gyr. \citet[HZ09]{hz09} analyzed the SFH of bright field stars using the StarFISH analysis software. The behaviour of the SFH predicted by \citetalias{hz09} does not seem to be consistent with that of the LMC’s cluster formation history because it predicts metallicities typically 0.2-0.3 dex more metal-poor than the observed values. Alternatively, \citet{rubele} used a combination of the minimization code StarFISH together with a database of ``partial models'' based on the CMDs of LMC populations of various ages and metallicities, from data of the Vista Magellanic Cloud (VMC) project. Note in Fig. \ref{f:fig20} that despite the smaller error bars involved due to the use of \citet{bressan} set of isochrones and although there is still a large dispersion in metallicity, there exist a clear tendency for the younger clusters to be more metal-rich than the intermediate age clusters. More precisely, clusters older than 1.2\,Gyr are significantly more metal-poor, with [Fe/H] $\leq$ -0.4. Because of the large dispersion shown in the diagram, there does not seem to be a sole model that can represent reasonably well the recent chemical evolution history in the LMC. More precisely, clusters older than $\sim$ 1.4 Gyr exhibit a slight tendency to decrease their metallicities with age, at odds with what \citetalias{pt98} predicted. Summing up, we believe that a model between those of \citetalias{pt98} and \citetalias{hz09} could probably be the most appropriate to represent the recent chemical evolution history in the LMC, but it is possible that an intrinsic metallicity spread exists at any given age, as found by \citep{parisi} in the Small Magellanic Cloud, and therefore a single monotonic chemical evolution model may not be appropriate. \\

\section{Summary}

We present for the first time $(T_1,C-T_1)$ CMDs for 17 poorly studied star clusters projected on the bar and on the inner disc and outer regions of the LMC. These objects are part of our ongoing project of generating a database of LMC star clusters homogeneously observed and studied by applying the same analysis procedure (P15; in preparation). As far as we know, none of these clusters has been previously studied. Ages and metallicities were determined from two different methods. In all cases, we compared the cleaned Washington CMDs with theoretical isochrones computed for the Washington system by the Padova group. The use of \citet{bressan} set of isochrones has largely reduced the metallicity associated errors resulting from the fittings. In some cases, we also estimated ages using the magnitude difference between the RGC and the MSTO, and derived metallicities by comparing the giant branches with standard calibrating clusters. We find a reasonably good agreement between the ages determined in these two different ways. Four clusters are found to be IACs (1.8-2.5 Gyr), whose metallicities range from -0.66 to -0.84. One cluster, SL\,263, is clearly young ($\sim$ 16 Myr), while the remaining 12 are aged between 0.32 and 0.89 Gyr and have metallicities in the range -0.50 $\leq$ Fe/H $\leq$ -0.19. By combining the current results with those of a sample of 231 additional clusters with ages and metallicities derived on a similar scale, we confirm a clear tendency for the younger clusters to be more metal-rich than the intermediate ones. The full cluster sample exhibit a significant dispersion of metallicities, whatever age is considered. This large metallicity dispersion does not seem to be related to the position of the clusters in the LMC. None of the chemical evolutionary models available in the literature satisfactorily represents the recent chemical evolution processes of the LMC observed clusters. \\

\section*{Acknowledgments}
We thank the staff and personnel at CTIO for hospitality and assistance during the observations. We especially thank the referee for his valuable comments and suggestions about the manuscript. We gratefully acknowledge financial support from the Argentinian institutions CONICET, FONCYT and SECYT (Universidad Nacional de C\'ordoba). T.P. gratefully acknowledges support provided by the Ministry of Economy, Development, and Tourism's Millennium Science Initiative through grant IC120009, awarded to The Millennium Institute of Astrophysics, MAS. D.G. gratefully acknowledges support from the Chilean BASAL Centro de Excelencia en Astrof\'isica y Tecnolog\'ias Afines (CATA) grant PFB-06/2007. This work is based on observations made at Cerro Tololo Inter-American Observatory, which is operated by AURA, Inc., under cooperative agreement with the National Science Foundation. This research has made use of the SIMBAD database, operated at CDS, Strasbourg, France; also the SAO/NASA Astrophysics data (ADS).

\bibliographystyle{mn2e}
\bibliography{paper}

\end{document}